\documentclass[12pt]{article}
\usepackage{amsmath,amssymb,amsbsy,amsthm}

\let\epsilon\varepsilon
\let\phi\varphi

\def\N{\mathbb N}

\newtheorem{theorem}{Theorem}

\newtheorem{corollary}{Corollary}
\newtheorem{claim}{Claim}

\begin{document}
\author{ Boris Ryabko, Daniil Ryabko \\ \{boris, daniil\}@ ryabko.net }
\title { Provably Secure Universal Steganographic Systems}
\date{}
\maketitle
\begin{abstract}
We propose a simple universal (that is, distribution--free) steganographic system
in which covertexts with and without hidden texts are statistically indistinguishable.
The stegosystem can be applied to any source generating i.i.d. covertexts with unknown distribution, and the hidden text is transmitted exactly, with zero probability of error.
Moreover, the proposed steganographic system has two important properties. First, the rate of transmission of hidden information approaches the Shannon entropy
of the covertext source as the  size of blocks used for hidden text encoding tends to infinity.
Second, if the size of the alphabet of the covertext source and its minentropy tend to infinity
then the  number of bits of hidden text per letter of covertext tends to $\log(n!)/n$ where $n$
is the (fixed) size  of blocks used for hidden text encoding. The proposed stegosystem uses randomization.
\end{abstract}

\section{Introduction}
The goal of steganography is as follows. Alice and Bob can exchange messages  of a certain
kind (called covertexts) over a public channel which is open to Eve. The covertexts can be, for example,
photographic images, videos, text emails and so on. Alice wants to pass some secret information
to Bob so that  Eve can not notice that any hidden information was passed. Thus,
  Alice should use the covertexts to hide the secret text. It is supposed that Alice and Bob
share a secret key. A classical illustration from  \cite{sy} states the problem
in terms of communication in a prison: Alice and Bob are prisoners who want
to concoct an escape plan passing each other messages which can be read by a ward.

Perhaps the first formal approach  to steganography was taken by Cachin \cite{ca, ca2}
who proposed a steganographic protocol in which, relying on the fact that
the probability distribution of covertexts is known, covertexts  with and without hidden
information are statistically indistinguishable.   In the same work a universal (distribution-free)
steganographic system was proposed, in which this property holds only asymptotically
with the size of the messages going to infinity, and which has exponential complexity of coding and decoding.
Distribution-free stegosystems are of particular practical importance, since in reality
covertexts can be graphical images, ICQ or email messages,  that is, sources for which
the distribution is not only unknown but perhaps can not be reasonably approximated.
Later a complexity--theoretic  approach for (distribution-free) steganography was developed in \cite{hla, le},
where stegosystems were proposed in which covertexts with and without hidden information
are indistinguishable in polynomial time.

We use the following model for steganography, mainly following \cite{ca2}.
It is assumed that Alice  has an access to an oracle which generates independent and identically
distributed covertexts
according to some fixed but unknown distribution $\mu$. Covertexts belong
to some (possibly infinite) alphabet $A$. Alice wants to use this source
for transmitting hidden messages. A hidden message is a sequence of letters from $B=\{0,1\}$
generated independently with equal probabilities of $0$ and $1$. We denote the source of hidden messages by $\omega$.
This is a commonly used  model for the source of secret messages since it
is assumed that secret messages are encrypted by Alice using a key shared only with Bob.
If Alice  uses the ideal Vernam cipher then the encrypted messages are indeed generated according
to the Bernoulli $1/2$ distribution, whereas if Alice uses modern block or stream ciphers then
the encrypted sequence ``looks like'' a sequence of random Bernoulli $1/2$  trials. Here
to ``look like'' means to be indistinguishable in polynomial time, or that the likeness is confirmed experimentally by
 statistical data, known for all widely used cyphers; see, e.g. \cite{rf,enc}.
The third party, Eve, is reading all messages passed from Alice to Bob and is trying
to determine whether secret messages are being passed in the covertexts or not.
Observe that if covertexts with and without hidden information have the same
probability distribution ($\mu$) then it is impossible to distinguish them.

In the universal system proposed in \cite{ca2}  the hiddentext sequence is divided into
blocks of a certain size $m$  each of which corresponds to
a block of length $n(m)$ of covertext letters from $A$. The distribution
of resulting covertext letters tends to the (unknown) distribution $\mu$
(of covertexts without hidden information) as $n$ tends to infinity. It is important to note
that the convergence is not uniform (on the set of all possible distributions $\mu$ with $A$ fixed), and
also the memory size of coder and decoder grows exponentially with $n$.

We propose a simple universal stegosystem for which
covertexts with and without hidden information have the same distribution
(and hence are statistically indistinguishable) for any size of the message.
The hidden text is transmitted correctly with probability 1.
Moreover, the proposed  system has two important properties.
First, the rate of transmission of hidden information approaches the Shannon entropy
of the covertext source as the  size $n$ of blocks used for hidden text encoding tends to infinity.
Second, if the size of the alphabet of the covertext source and its minentropy tend to infinity
then the  number of bits of hidden text per letter of covertext tends to $\log(n!)/n$ where $n$
is the (fixed) size  of blocks used for hidden text encoding.
The latter property is, in particular, an advantage as compared to the complexity--theory based stegosystems proposed
in~\cite{hla, le, al} for which the rate of hidden text transmission is no more than a constant per covertext letter. 
We note  that it is also possible to use the proposed stegosystems
for open-key steganography in a standard way. 
                                            
The paper is organized as follows.
In Section~\ref{sec:nonr} a simple stegosystem which does not use randomization is proposed;
for this system the number of bits of hidden text per letter of covertext tends to $1/2$ if
the size of the alphabet of the covertext source and its minentropy tend to infinity.
This system also illustrates the main ideas used in Section~\ref{sec:main}, where
the general (randomized) stegosystem is proposed which has the mentioned asymptotic
properties of the rates of hidden text transmission.
In Section~\ref{sec:disc} we discuss possible extensions of the proposed steganographic systems
and outline some potentially interesting open problems.
In   particular, we discuss issues concerning stegosystems based on a common set of data
and open--key steganography.

\section{A simple non-randomized universal \\ stegosystem}\label{sec:nonr}
In this section we present a very simple stegosystem which demonstrates
the main ideas used in the general stegosystem which we develop in the next section.
The stegosystem described in this section does not use randomization.

The notation is as follows. The source $\mu$ draws i.i.d. (covertext) letters from an alphabet $A$.
The source $\omega$ draws i.i.d. (hidden, or secret) equiprobable letters 
from the alphabet  $B=\{0,1\}$.
Finite groups of (covertext, hidden, secret) letters are sometimes called (covertext, hidden, secret) words.
Elements  of $A$ ($B$) are usually denoted by $x$ ($y$).

First consider an example. Consider a situation in which not only the
secret letters  are drawn (using $\omega$) from a  binary alphabet, but also the source of covertexts $\mu$ generates
symbols from the alphabet $A=\{a,b\}$ (not necessarily with equal probabilities).
Suppose that Alice has to transmit the sequence $y^*=y_1y_2\dots$ generated according
to $\omega$ and let there be given a covertext sequence $x^*=x_1x_2\dots$ generated by $\mu$.
For example, let
\begin{equation}\label{eq:ex}
y^*=01100\dots, \ \ x^*=aababaaaabbaaaaabb\dots.
\end{equation}
The sequences $x^*$ and $y^*$ are encoded in a new sequence $X$
(to be transmitted to Bob) such that $y^*$ is uniquely determined
by $X$ and the distribution of $X$ is the same as the distribution
of $x^*$ (that is, $\mu$; in other words, $X$ and $x^*$ are
statistically indistinguishable).

The encoding is carried out in two steps. First let us group all symbols of $x^*$ into pairs, and
denote
$$
 aa=u,\ bb=u,\ ab=v_0,\ ba=v_1.
$$
In our example, the sequence~(\ref{eq:ex}) is represented as
$$
x^*=aa\,ba\,ba\,aa\,ab\,ba\,aa\,aa\,bb \dots=uv_1v_1uv_0v_1uuu\dots
$$
Then $X$ is acquired from $x^*$ as follows: all pairs corresponding to $u$ are left unchanged,
while all pairs corresponding to $v_k$ are transformed to pairs corresponding to $v_{y_1}v_{y_2}v_{y_3}\dots$;
in our example
$$
 X= aa \, ab \, ba \, aa \, ba \, ab \,  aa \,  aa \,  bb ... .
$$
Decoding is obvious: Bob groups the symbols of $X$ into pairs, ignores all occurrences of $aa$ and $bb$
and changes $ab$ to $0$ and $ba$ to $1$.

The properties of the described stegosystem, which we call $St_2$, are summarized in the following (nearly obvious) statement.
\begin{claim}
Suppose that a source $\mu$ generates i.i.d. random variables
taking values in $A=\{a,b\}$ and let this source be used for encoding
secret messages consisting of a sequence of i.i.d. equiprobable binary
symbols using the method $St_2$.
Then the sequence of symbols output by the stegosystem
obeys the same distribution $\mu$ as the input sequence.
\end{claim}
We will not give the (obvious) proof of this claim since it is a simple corollary
of Theorem~\ref{th:nonr} below.

It is interesting to note that a similar construction was used by von Neumann in his
method for obtaining a sequence of equiprobable binary symbols (see \cite{fn,el}) from a sequence of independent flips of a biased coin.
His method, as well as the just described stegosystem, was based on the fact that the probabilities of $ab$ and $ba$ are equal.

Next we consider the generalisation of the described stegosystem to the case of any alphabet $A$ (such that  $|A|>1$).
To do this we fix some  total ordering on the set $A$.
As before, Alice has to transmit a sequence $y^*=y_1y_2\dots$ generated by the source $\omega$ of i.i.d. equiprobable
binary letters and let there be given a sequence $x^*=x_1x_2\dots$ of covertext letters  generated
i.i.d. according to a distribution $\mu$ on  $A$.  Again we transform
the sequences $y^*$ and $x^*$ into a new sequence $X$ which obeys  the same distribution as $x^*$.
As before we break $x^*$ into blocks of length $2$. If a block $x_{2i-1}x_{2i}$ has the form $aa$ for some $a\in A$ then
it is left unchanged. Otherwise let the block  $x_{2i-1}x_{2i}$ be $ab$ for $a,b\in A$ and suppose $a<b$;
if the current symbol $y_k$ is $0$ then the block $ab$ is included in $X$, and if $y_k=1$ then $ba$ is included in $X$.
If $a>b$ then encode in the opposite way. To decode, the sequence is broken into pairs of symbols, all pairs of the form $aa$
are ignored and a pair of the form $ab$ is decoded as $0$ if $a<b$ and as $1$ otherwise.
Denote this stegosystem by $St_2(A)$.

\begin{theorem}\label{th:nonr}
Suppose that a source $\mu$    generates i.i.d. random variables
taking values in  some alphabet $A$. Let this source be used for encoding
secret messages  consisting in a  sequence of i.i.d. equiprobable binary
symbols, using the method $St_2(A)$.
Then the sequence of symbols output by the stegosystem
obeys the same distribution $\mu$ as the input sequence and the number
of letters of hidden text transmitted per letter of covertext is $ \frac{1}{2}( 1 -  \sum_{a \in A}  \mu(a)^2)$.
\end{theorem}
\begin{proof}
Fix some     $\alpha,  \beta \in A $
and $k\in\N $. We will show that
$$ p( X_{2k -1}X_{2k} =  \alpha  \beta ) \, = \, \mu (\alpha  \beta), $$
where $p$ is the probability distribution of the output sequence.
Suppose $\alpha<\beta$. Decomposing the probability on the left we get
\begin{multline*}
 p( X_{2k -1}X_{2k}
=  \alpha  \beta )  =
 \omega(y_{k} = 0) (\mu (\alpha  \beta)    +    \mu (  \beta \alpha) )\\
 = \frac{1}{2}(\mu(\alpha\beta)+\mu(\alpha\beta))= \mu (\alpha  \beta)   .
\end{multline*}
The case $\beta<\alpha$ is analogous, and the case $\beta=\alpha$ is trivial.
The second statement is obtained by calculating the probability  that  letters in the block coincide.
\end{proof}
Note that in practice when the covertexts are, for example, graphical files, each covertext is
practically unique (the alphabet $A$ is potentially infinite) so that the number of covertext letters (files)
per one hidden bit is approximately~2.

\section{General construction of  a universal \\ stegosystem}\label{sec:main}
In this section we consider the general construction of universal stegosystem which has
the desired asymptotic properties.
As before, Alice needs to  transmit a sequence $y^*=y_1y_2\dots$ of secret binary messages drawn
by an i.i.d. source $\omega$ with equal probabilities of $0$ and $1$, and let there
be given a sequence of covertexts $x^*=x_1x_2\dots$ drawn i.i.d. by a source $\mu$ from an alphabet $A$.
First we break the sequence $x^*$ into blocks of $n$ symbols each, where  $n>1$ is a parameter.
Each block will be used to transmit several symbols from  $y^*$  (for example, in the
previously constructed stegosystem $St_2(A)$ each block of length 2 was used to transmit 1 or 0 symbols).
However, in the general case a problem arises which was not present in the construction of $St_2(A)$.
Namely, we have to align the lengths of the blocks of symbols from $x^*$ and from $y^*$, and
for this we will need randomization. The problem is that
the probabilities of blocks from $y^*$ are divisible by powers of $2$, which is not necessarily the
case with blocks from $x^*$.

We now present a formal description. Let $u$ denote the first $n$ symbols of $x^*$: $u=x_1\dots x_n$, and
let $\nu_u(a)$ be the number of occurrences of the symbol $a$ in $u$. Define the set $S_u$ as consisting
of all words of length $n$ in which the frequency of each letter $a\in A$ is the same as in $u$:
$$
S_u=\{v\in A^n: \forall a\in A\ \nu_v(a)=\nu_u(a)\}.
$$
Observe   that the $\mu$-probabilities of all members of $S_u$ are
equal. Let there be given some ordering on the set $S_u$ (for
example, lexicographical) which is known to both Alice and Bob
(and to anyone else) and let $S_u=\{s_0,s_1,\dots $ $,s_{|S_u|-1}\}$
with this ordering.

Denote $m =\lfloor{\rm log}_2|S_u|\rfloor$, where $\lfloor y\rfloor$ stands for the largest integer not greater than $y$.
Consider the binary expansion of $|S_u|$:

$$
|S_u|=(\alpha_{m},\alpha_{m-1},\ldots,\alpha_{0}) ,
$$
   where $\alpha_{m}=1$, $\alpha_j\in\{0,1\}$ , $m > j \geq0$.
In other words,
$$
  |S_u| =   2^m + \alpha_{m-1} 2^{m-1} +  \alpha_{m-2} 2^{m-2}+ ... + \alpha_0. 
$$
Define a random variable $\Delta$ as taking each value $i\in\{0,1,\dots,m\}$
with probability $ \alpha_{i} 2^{i} / |S_u|:$
\begin{equation}\label{del}
p(\Delta = i)  =    \alpha_{i} 2^{i} / |S_u|  .
\end{equation}
Alice, having read $u$, generates a value of the random variable $\Delta$, say $d$,
and then reads $d$ symbols from $y^*$.
Consider the word $r^*$ represented by these symbols as an integer which we denote by $r$.
Then we encode the word $r^*$ (that is, $d$ bits of $y^*$) by the word $s_\tau$
from the set $S_u$, where
$$
   \tau = \sum_{l =d+1}^{m}  \alpha_{l} 2^{l}  + r   .
$$
(In other words, the word $s_\tau$ is being output by the coder.)

Then Alice reads the next $n$--bit word, and so on. Denote the
constructed stegosystem by $St_n(A)$.

To decode the received sequence Bob breaks it into blocks of length $n$ and
repeats all the steps in the reversed order: by the current word $u$ he obtains $S_u$ and $\tau$,
then $d$ (clearly $d$ is uniquely defined by $\tau$), $r$
 and $r^*$; that is,
he finds $|r^*|$ next symbols of the secret sequence $y^*$.

Consider an example which illustrates all the steps of the calculation.
Let  $A= \{ a,b,c \}, \, n = 3, \,
 u = bac .$   Then  $S_u = \{   abc, acb, bac, bca, cab, cba \},   \, |S_u| = 6, m = 2,
\alpha_2 = 1, \alpha_1 = 1, \alpha_0 = 0. $
Let the sequence of secret messages be
 $0110... ,$ that is,   $y^* = 0110... \, .$
Suppose the value of $\Delta$ generated by Alice is  1.
Then she reads one symbol of  $y^* $ (in this case  0) and calculates
 $r = 0, r^* = 0 , \tau =  2^2 + 0 = 4 $ and finds the codeblock $s_4 = cab.$
To decode the message, Bob from the block $cab$ calculates $\tau = 4, r = 0, r^* = 0$ and finds the
next symbol of the secret sequence --- 0.

\begin{theorem}\label{th:main}
Let a source $\mu$ be given, which  generates i.i.d. random variables
taking values in some alphabet $A$. Let this source be used for encoding
secret messages  consisting of a sequence of i.i.d. equiprobable binary
symbols using the described method $St_n(A)$ with $n>1$.
Then
\begin{itemize}
\item[(i)]
the sequence of symbols output by the stegosystem
obeys the same distribution $\mu$ as the input sequence,
\item[(ii)] the average number of secret symbols per
covertext  ($L_n$) satisfies the following inequality
\begin{equation}\label{th}
L_n \geq  \frac{1}{n}  \left( \sum_{u \in A^n}  \mu(u) \log  \frac{n!}{\prod_{a \in A  } \nu_u (a)!   } -  2\right),
\end{equation}   where                            $\mu(u)$ is the $\mu$-probability of the word $u$
and $\nu_u(a)$ is the number of occurrences of the letter $a$ in the word  $u$.
\end{itemize}
\end{theorem}

\begin{proof}
To prove the first statement it is sufficient to show that for any covertext word $u$ of length $n$
its probability of occurrence in the output sequence is $1/|S_u|$.
This follows from~(\ref{del})  and the fact that letters in $y^*$ are independent and equiprobable.

The second statement can be obtained by direct calculation of the average number
of symbols from $y^*$ encoded by one block. Indeed, from~(\ref{del})  we find that
 for each covertext word $u$ the expected number of transmitted symbols is $\frac{1}{|S_u|}\sum_{l=1}^m l\alpha_l2^l \ge |S_u|-2$, where $m =\lfloor{\rm log}_2|S_u|\rfloor$,
and for each word $u$ we have $|S_u|=\frac{n!}{\prod_{a \in A  } \nu_u (a)!   } $.
\end{proof}

Let us now consider the asymptotic behaviour of $L_n$ when $n\rightarrow\infty$.
\begin{corollary}
If the alphabet $A$ is finite
then the average number of hidden symbols per letter $L_n$ goes to the
Shannon entropy $h(\mu)$ of the source $\mu$ as $n$ goes to infinity; here
by definition  $h(\mu)=  - \sum_{a \in A}  \mu(a) \log  \mu(a)$.
\end{corollary}
\begin{proof}
This statement follows from a  well-known  fact of Information
Theory which states that for each  $\delta > 0$ and $n \rightarrow
\infty$ the following inequality holds with probability 1
$$ h(\mu)- \delta < \log |S_u| / n < h(\mu) + \delta, $$
see, e.g.   \cite{ga}.
\end{proof}

In many real stegosystems the alphabet $A$ is huge (it can consist, for example, of
all possible digital photographs of given file format, or of  all possible e-mail messages).
In such a case it is interesting to consider the asymptotic behaviour
of $L_n$ with fixed $n$ when the alphabet size $|A|$ goes to infinity.
For this we need to define the so-called  min-entropy of the source $\mu$:
\begin{equation}\label{me}
H_\infty(\mu) = \min_{a \in A} \{  - \, \log \mu (a) \}  \,  .
\end{equation}

\begin{corollary}
Assume the conditions of Theorem~\ref{th:main} and fix the block length $n>1$.
If $|A|\rightarrow\infty$ so that $ H_\infty(\mu)
 \rightarrow \infty$ then   $L_n$ tends to  $ \ (log (n!) - O(1) ) / n  $.
\end{corollary}
This statement simply follows from the fact that the number
of different permutations of $n$ elements is $n!$.

Next we briefly consider the resource complexity of the stegosystem  $St_n(A).$
To store all possible words from the set $S_u$ would require memory of order  $2^n \log |A|$ bits,
which is practically unacceptable for large $n$.
However, if we use the algorithm for fast enumeration from \cite{r}, then
we can find the index of a block $s_\tau$ given $\tau$ (encoding) and
vice versa (decoding) using  $ O( \log^{const} n)$
operations per symbol and  $O(n  \log^{3} n) $ bits of memory.

\section{Discussion}\label{sec:disc}
We have proposed two  stegosystems (with and without randomization)
for which the output sequence of covertexts with hidden information
is statistically indistinguishable from a sequence of covertexts without
hidden information.
The proposed stegosystems rely heavily
on the assumption that the oracle generates independent and
identically distributed covertexts. This is perhaps a reasonable
assumption if covertexts are graphical images of a certain kind,
but if, for example, we want to use just one image to transmit (a
large portion of) a secret text then our covertexts are parts of
the image, which are clearly not i.i.d. How to extend the ideas
developed in this work to the case of non-i.i.d. covertexts is
perhaps the main open question.

However,  the main idea that was used in the proposed stegosystems is that
for any block of covertexts it is possible to find several other
blocks which have the same probability as the original one; then
hidden information can be encoded in the number of a block in this group.
This idea can be extended to the case of non-independent covertexts.
Indeed, suppose that on the current step of transmission we known that 
some covertexts have equal probabilities to appear as the next generated covertext. 
That is, among the conditional (given the current history) probabilities of covertexts there
are several groups of equal probabilities. Then, if the probability of the next generated covertext 
belongs to one of these groups,  we can use this covertext (possibly replacing it 
with another one which has the same probability) for encoding several next bits of hiddentext in the
same fashion  as it is done in $St_n(A)$. The same applies to blocks of covertexts. Indeed the only feature
of independently and identically distributed covertexts that we used was that all permutations within a block of size $n$ have equal probabilities.
So the next step is to identify equal conditional probability groups in sources of non-i.i.d. covertexts.


\end{document}